\documentstyle[multicol,aps,epsf]{revtex} 
\tightenlines
\begin{document}
\title{ Electric field dependent structural and  vibrational properties of the Si(100)-H(2$\times$1) surface  and its implications for STM induced hydrogen desorption}
\author{K. Stokbro }
\address{Mikroelektronik Centret, Danmarks Tekniske Universitet, 
Bygning 345\o , DK-2800 Lyngby, Denmark.}

\date{\today}
\maketitle

\begin{abstract}
We report a first principles study of the structure and the vibrational properties of the
Si(100)-H(2$\times$1) surface in an electric field.  The
calculated vibrational parameters are used to model the
vibrational modes in the presence of the electric field corresponding to a
 realistic STM tip-surface geometry. We find that
local one-phonon  excitations  have short lifetimes (10 ps
at room temperature)
due to incoherent lateral diffusion, while     diffusion of
local multi-phonon excitations are suppressed due to  anharmonic frequency
shifts and have
much longer lifetimes (10 ns at room temperature). We
calculate the implications for current induced desorption of H using a
recently developed first principles model of  electron inelastic
scattering. The calculations show that  inelastic scattering events
with energy transfer $n \hbar \omega$, where $n>1$, play an important
role in the desorption process.  

\end{abstract}
\begin{multicols}{2}
\narrowtext
\section{INTRODUCTION}
STM induced desorption of
hydrogen(H) from the
monohydride Si(100) surface  offers the possibility of lithography with  atomic resolution\cite{BeHiChBe90,LyAv90,LyAbShWaTu94}.  Investigations of the
desorption
mechanism\cite{ShWaAbTuLyAvWa95,AdMaSw96,AvWaRoShAbTuLy96,ScRaEnHaKo96,AvWaRoAkNoShAbLy96,ShAv97,FoKaLyAv98,StThSaQuHuMuGr98,StHuThXi98,ThSaNaSt99}
have established  the dependence of the desorption rate on the  bias voltage,
tunnel current and H isotope. 
At high positive biases, $V_{b}>4$ V, the experimental results are consistent with 
electron induced desorption due to direct excitation of the Si-H
bond by a single
electron\cite{ShWaAbTuLyAvWa95,AdMaSw96,AvWaRoShAbTuLy96}. At negative
 and low positive biases the desorption rates show
power-law dependencies on the electron
current\cite{ShWaAbTuLyAvWa95,StThSaQuHuMuGr98} consistent with
a multi-electron process\cite{ShWaAbTuLyAvWa95,GaPeLu92,WaNeAv93}, and  the measured
desorption rates are in quantitative agreement
 with first-principle calculations\cite{StThSaQuHuMuGr98,StHuThXi98}. 

The desorption by  multi-electron scattering is only possible because the H  stretch
frequency has a long lifetime. The long lifetime is a result of a
vibrational quantum  too low for coupling with  electron-hole
excitations, while  well above  the Si  phonon
spectrum, and can thus  only transfer  energy to the substrate via a
multi-phonon process. At room
temperature experimental estimates of the lifetime due to this process are
$\tau\approx10$ ns\cite{GuLiHi95,FoKaLyAv98}. However, a local  excitation is
not an eigenmode and will decay into a H surface phonon by a
coherent process. This decay is  several orders of magnitude
faster than multi-phonon energy relaxation, and must
therefore  be included in the theoretical models.
 It has been
proposed  by Persson and Avouris\cite{PeAv95,PeAv97} that the vibrational Stark shifts due to the 
 electric field from the tip   can localize  vibrational
modes below the tip.  The localized modes may still transfer
energy laterally by  incoherent diffusion (the F\"{o}rster mechanism)
but  it was found that   this decay channel is also
reduced by the Stark shifts.   However, the work  assumed that the
electric field  from the STM tip is localized on a
single H atom below the tip, and this is not the case for realistic tip
geometries. 

 In this work we present results for the vibrational properties of the
Si(100)-H(2$\times$1) surface in the presence of the electric field
from a more realistic model for the STM tip. The tip is described by
sphere of radius, $R_t=500$~\AA , with a protrusion of atomic dimensions,
and we determine  the electric field by  solving  the
Poisson's equation numerically. To  obtain the effect on the H
vibrations we set up a phonon Hamiltonian with  parameters obtained
from a  first principles calculation of the
vibrational properties of the Si(100)-H(2$\times$1) surface in  an
external electric field. We find that the electric field  does  give rise to a localized
vibrational state below the tip; however, its lifetime is very
short (10 ps) due to
incoherent exciton motion. However, we find that the
anharmonicity of the Si-H bond potential reduces the lateral energy
transfer of higher excited 
excitations($n>1$) of the Si-H bond.  We present first principle calculations of the desorption rate taking this
effect into account, and find that two phonon excitations play an
important role in the desorption process. 

The organization of the paper is the following. In Section~\ref{sec:firstp0} we
describe the first principles method which  in section~\ref{sec:firstp} is used to calculate 
the zero field atomic structure  and Si-H stretch frequencies of the Si(100)-H(2$\times$1)
 surface.   The electric  field
dependence of the frequencies is calculated in Section~\ref{sec:field}.
 In Section~\ref{sec:stark} we introduce a simple dipole-dipole interaction 
 model for  the Si-H stretch phonon
band, and uses it to find localized vibrational states in the presence of 
 electric fields from  different STM tip geometries. In
Section~\ref{sec:decay} we calculate the lifetimes of the localized
states due to incoherent lateral diffusion. In 
section~\ref{sec:des} the lifetimes are used to model STM induced
desorption. Section~\ref{sec:sum} summaries the results.

\section{Structure and vibrational properties of the
Si(100)-H(2$\times$1) surface.}
\label{sec:firstp0}
In this section we calculate the vibrational frequencies and the
dipole-dipole coupling matrix elements of the H vibrations on the
Si(100)-H(2$\times$1) surface. In subsection~A we present calculations for the unperturbed Si(100)-H(2$\times$1) surface, and the
vibrational and structural shifts due to an external planar field are
calculated  in subsection~B.

The first principles calculations are based on density functional
theory\cite{HoKo64,KoSh65} within the  Generalized Gradient
Approximation (GGA)\cite{PeWa91} for the exchange-correlation
energy. Since we  only consider filled shell systems, the calculations
are all non-spin-polarized.  Ultra-soft pseudo
potentials\cite{Va90} constructed from a scalar-relativistic
all-electron calculation are used to describe H and
silicon(Si)\cite{pseudo}. The wave functions are expanded in a plane-wave
basis set with a kinetic-energy cutoff of 20 Ry, and with this choice
absolute energies are converged better than $0.5$ mRy/atom.

 With this approach we find a Si lattice constant of 5.47(5.43\cite{CRC94})\AA\
and bulk modulus of 0.89(0.97\cite{CRC94}) Mbar (parentheses show
experimental values). For the  H$_2$ 
molecule we obtain a bond length of
0.754(0.741\cite{CRC94})\AA, a binding energy (including zero point motion) of
4.22(4.52\cite{CRC94}) eV and a vibrational frequency of
4404(4399\cite{CRC94})cm$^{-1}$. Generally the comparison with
experiment is excellent, and similar theoretical values have been
found in other studies using the GGA\cite{PeWa91,CoBaRe94}.

\subsection{ Zero field properties }
\label{sec:firstp}
To model the monohydride Si(100) surface at zero field we use a
(2$\times$1) slab with 12 Si atoms and 6 H atoms.  The atoms at
the bottom surface are bulk like, and their dangling bonds are  saturated
with  H atoms. The
two surfaces are separated by a 7.5~\AA\ vacuum region, and we use the
dipole correction\cite{NeSc92} in order to describe the different
workfunction  of the two
surfaces.  The surface is insulating and we use 2  $k$-points in the
irreducible part of the Brillouin-zone (BZ) for the BZ integrations. 
  Test calculations with more dense meshes show that BZ
integration errors are negligible small.

 Figure~1 shows the atomic structure  after relaxation of the H
atoms and the 4 upper Si layers.  Positions of the H
atoms and the two upper Si
layers compare well with other studies\cite{JiWh95,KrHaNo95,RaCa96}.
For the  third and fourth
layer Si atoms  we find a small asymmetric relaxation, and to our
knowledge such relaxations have not been included in previous studies.

\begin{figure}
\begin{center}
\leavevmode
\epsfxsize=84mm
\epsffile{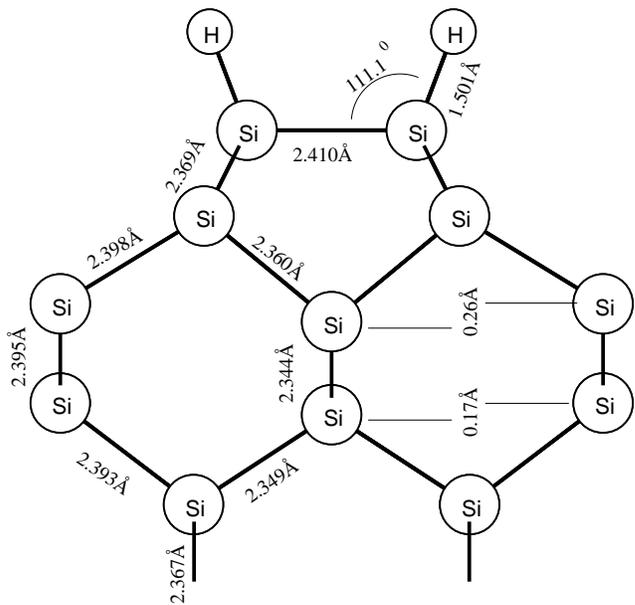}
\end{center}
\caption{Calculated atomic structure of the  Si(100)-H(2$\times$1) surface. } 
\end{figure}

To obtain the  dynamical matrix of the Si-H stretch frequency we make 
 H displacements of $\pm 0.07, 0.14,\ldots ,0.35$~\AA\ in the Si-H bond
direction and  fit a sixth order
polynomial to the data points. Since the Si-H stretch frequencies 
are four times higher than Si bulk frequencies, the
Si substrate acts like  a solid wall, and we therefore  use the H mass ${\rm M_H}$ when
calculating frequencies from the dynamical matrix. 
 With this approach we have calculated 
 the  frequency of the
symmetric stretch $\omega_{\rm s}$ and the asymmetric stretch
$\omega_{\rm a}$ at three high symmetry points in the surface Brillouin
zone\cite{cellsize}.  In Table~1 the results are listed together  with   $\Gamma$
point frequencies(parenthesis) obtained by  infrared
 spectroscopy\cite{ChRa84} and the comparison  between theory and
 experiment is excellent, especially we note that theory correctly
 predicts the  splitting, $\omega_{\rm s}-\omega_{\rm a}=11$~cm$^{-1}$.

\begin{table}
\caption{Surface phonon frequencies in cm$^{-1}$ for the 
 H symmetric stretch $\omega_{\rm s}$ and asymmetric stretch
$\omega_{\rm a}$ at the $\Gamma$, J and J' point in the surface Brillouin
zone. Experimental $\Gamma$ point  frequencies are shown in
parenthesis. Bracketed J and J' 
point  frequencies are obtained from a dipole-dipole
interaction model, with parameters fitted to the two $\Gamma$ point
frequencies (see Section~\protect\ref{sec:stark}). } 
\begin{tabular}{llll}
\multicolumn{1}{l}{( $2\pi/a_{100}$)} & \multicolumn{1}{c}{$\Gamma=(0,0)$} & 
\multicolumn{1}{c}{$J=(0,1/4)$} &\multicolumn{1}{c}{$J'=(1/2,0)$}  \\
\tableline
$\omega_{\rm s}$ (cm$^{-1}$)    & 2082 (2099$^a$) & 2075[2075] & 2074[2071] \\
$\omega_{\rm a}$ (cm$^{-1}$)  & 2071 (2088$^a$) & 2074[2073] & 2074[2069] \\
\end{tabular}
$^a$Reference~\protect\cite{ChRa84}
\end{table}

We next investigate the bonding potential of the H atoms.  In Fig.~2
the solid circles show the hydrogen energy, $E_{\rm H}$,  for the
 Si-H bond lengths, $z$,  used  in the
calculation of the dynamical matrix. 
The data are accurately described by a
Morse potential
\begin{equation}
E_{\rm H}(z)= E_{\rm d} (e^{-2
\alpha (z-z_0)}-2e^{-\alpha (z-z_0)}),
\end{equation}
 and from a least-squares fit we obtain the  frequency
$\omega_0=0.26$ eV($\alpha=1.57$~\AA$^{-1}$), equilibrium bond length
$z_0=1.50$~\AA , and desorption
barrier $E_{\rm d} =3.4$ eV. The extrapolated desorption energy
coincides with the surface energy
without a H atom plus   the energy  of a
spin-polarized H atom. The triangles 
in Fig.~2 show the total energy for large values of $z-z_0$. When the interaction between the H atom and the
surface becomes weak the electrons start to spin-polarize and the data
points in Fig.~2
show that spin polarization effects become important 
for $z>2.5$~\AA . 

The inset shows the change in the surface dipole, $\Delta p$ as a
function of $z$ (the positive direction is from Si to H). The  surface dipole
increases almost linearly with $z$ and the
  dynamic dipole moment is $\gamma=\partial p/\partial z = 0.6 $
Debye/\AA\ (=0.13 e). Modelling  the  surface dipole by  an effective charge
$e^*$
on the H atom and its image charge -$e^*$\cite{image}, we find  $e^*\approx
-0.07e$. The sign of this  charge transfer from Si to H is in
accordance with a
 higher electronegativity of H relative to Si\cite{AkNoLoAv97}.

\begin{figure}
\begin{center}
\leavevmode
\epsfxsize=84mm
\epsffile{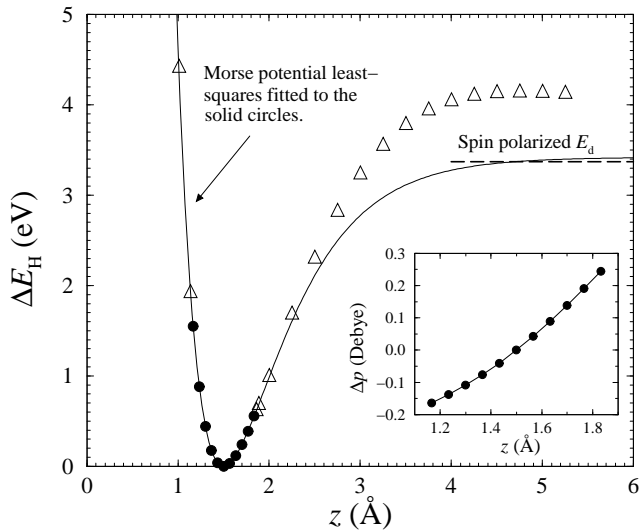}
\end{center}
\caption{Filled circles and open triangles show the change in the
total energy $\Delta E_{\rm H}$ as a function of the
 Si-H bond length $z$, and  the solid line shows a Morse
potential fit to the filled circles data points. The horizontal dashed line shows the
desorption energy, $E_d$, as obtained from a spin-polarized H atom and
the surface without a H atom. To obtain the dynamical matrix
we fit a sixth order polynomial to the filled circles data
points. Inset shows the change in the dipole-moment $\Delta p$ as
function of $z$.}
\end{figure}

\subsection{Electric field dependent properties}
\label{sec:field}
To model the surface in a planar external field we use
 a  (2$\times$1) slab  with 24 Si
atoms, 2 H atoms, and a vacuum region of 10~\AA\ and the external 
 field  is modelled using the method of Ref.~\cite{NeSc92}. The Si atoms at the
back side of the slab are not passivated by H atoms, and dangling bonds on these atoms can donate free electrons and holes. In this way we take into
account the effect of mobile carriers\cite{KrHaGrNo96}. Other
 computational details are identical to those for the zero field calculation.

\begin{figure}
\begin{center}
\leavevmode
\epsfxsize=84mm
\epsffile{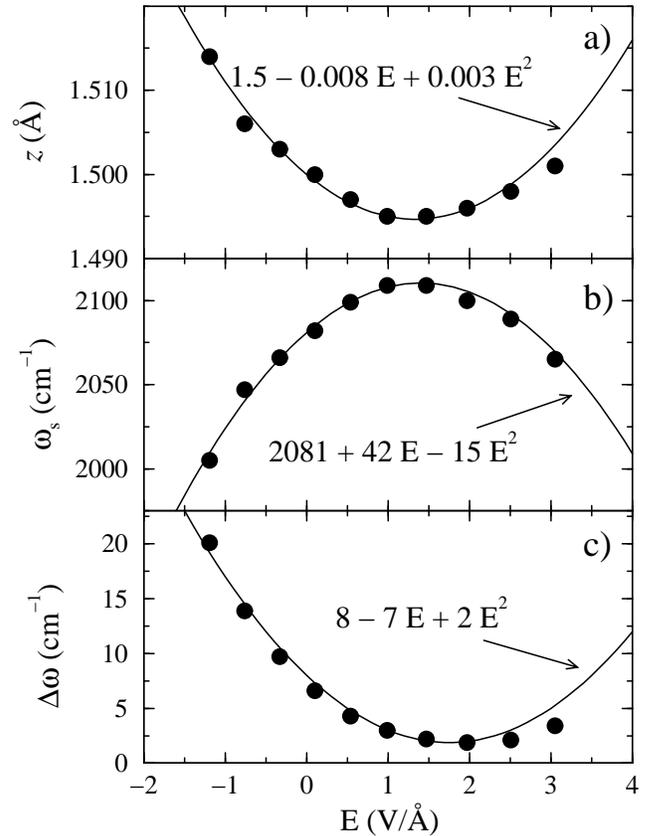}
\end{center}
\caption{The field dependence of (a) the equilibrium Si-H bond length
$z_0$, (b) the $\Gamma$ point symmetric Si-H stretch frequency
$\omega_{\rm s}$, (c) the $\Gamma$ point symmetric--asymmetric splitting
$\Delta \omega = \omega_{\rm s} - \omega_{\rm a}$. The solid lines show second
order polynomials  obtained by least-squares fit to the data.}
\end{figure}

Curves in Fig.~3 show the field dependence of the equilibrium Si-H bond length $z$,  the $\Gamma$ point symmetric Si-H stretch frequency
$\omega_{\rm s}$ and the $\Gamma$ point symmetric--asymmetric splitting
$\Delta \omega = \omega_{\rm s} - \omega_{\rm a}$.
We first notice that all three quantities  have an extremum at $\approx 1.5$
V/\AA . This  behaviour can be described by 
a simple Si-H tight-binding model with a field dependent H on-site
element\cite{PeAv95}. The extrema occurs at the field where the
H and Si on-site levels are in resonance, since at
 resonance the Si-H bond is strongest, and therefore the bond length minimal and the
vibrational frequency maximal. Furthermore, at resonance the H dynamic
dipole vanishes, and therefore also the  part of $\Delta \omega$ 
caused by H-H  dipole interactions.

The three solid lines in Fig.~3 show  second order polynomials
obtained by  least squares fit to the data. The
interpolated  zero field values of
$z_0$ and $\omega_{\rm s}$ agrees  exactly with those obtained
in section~\ref{sec:firstp}, while the interpolated
$\Delta \omega$  zero field value is slightly off.
Taking into account the quite different slabs used for the two
calculations we find the agreement fully satisfactory, and note that
the  difference  can  be
taken as a measure of the accuracy of the approach.

Recently, the electric field dependent properties of the  H/Si(111)(1$\times$1)
surface were calculated by Akpati {\it et~al.}\cite{AkNoLoAv97}, and they found 
Stark shifts  $\approx 30$ percent larger than in
the present calculation, and the extremum in bond length and frequency 
appears for a field of 1~V/\AA . The agreement  with the present calculation
 seems reasonable, bearing in mind
that the Stark shifts are for different crystallographic surface directions. 
However, part of the difference might be due to the use of a cluster
geometry and the local spin density approximation in
Ref.~\cite{AkNoLoAv97}. The present study is based on    a slab
geometry and the GGA. We expect that the thick slab geometry better
describes the electric field induced polarization of the surface.  

\section{STM induced Stark localization}
\label{sec:stark}
In this section we will model the collective modes of
the Si-H stretch vibrations by a set of local oscillators
interacting through dipole forces\cite{PeRy81,elastic}, and use this model to calculate 
the Stark localization in the external  electric field from a STM
tip. In Fig.~4 is
shown the lattice sites of the oscillators, corresponding to the
positions of the H atoms in the (2$\times$1) cell. 
Each oscillator is described by a local frequency 
$\omega_i$ and a dynamic dipole moment $\gamma_i$. The Hamiltonian of
the system is given by
\begin{equation}
H_{ij}=  \hbar \omega_i \delta_{ij} + \frac{\chi_i \chi_j}{ |{\bf
r}_i-{\bf r}_j|^3}(1-\delta_{ij}), 
\label{eq:hamil}
\end{equation}
where $r_i$ is the position of oscillator $i$, and $\chi_i
=\sqrt{\hbar/2{\rm M_H}\omega_i}\gamma_i$. 

\begin{figure}
\begin{center}
\leavevmode
\epsfxsize=84mm
\epsffile{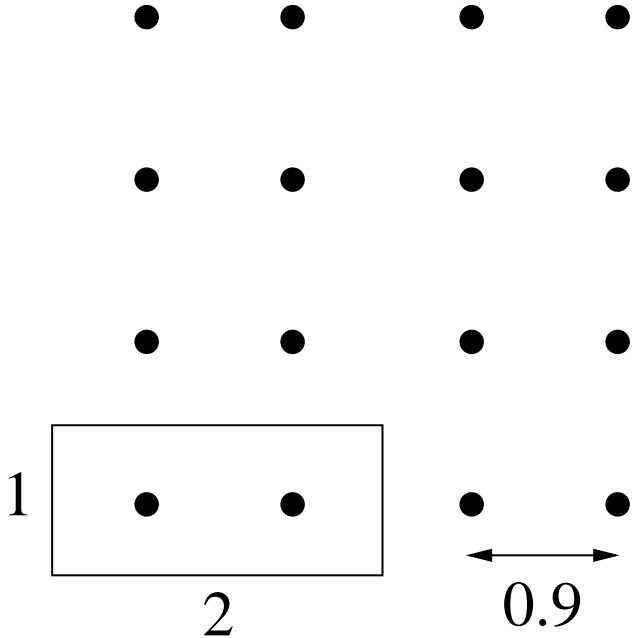}
\end{center}
\caption{The  (2$\times$1) surface lattice of  H atoms, in units of the
surface lattice constant $a_{100}=3.87$~\AA .}
\end{figure}

We first consider the zero field case of  identical oscillators with parameters $\gamma_0$, $\omega_0$. We  find the dispersion from a numerical Fourier transform, and 
at the  $\Gamma$ point  we obtain the Hamiltonian
\begin{equation}
H(\Gamma) = \left(
\begin{array}{cc}
\omega_0 + 4.05 V_0   & 5.12 V_0 \\
5.12 V_0   & \omega_0 + 4.05 V_0 \\
\end{array}
\right),
\end{equation}
where $V_0 = (\gamma_0^2)/({\rm M_H} \omega_0 a_{100}^3)$  and
$a_{100}=3.87$~\AA\  is the surface lattice constant. The two eigenmodes are
 $\omega_{\rm s}(\Gamma) = \omega_0+9.17 V_0$, and $\omega_{\rm a}(\Gamma)=\omega_0 -1.07
V_0$. Using the  calculated values of $\gamma_0$ and $\omega_0$ from 
section~2A, we 
obtain $V_0=0.07$ meV and thereby  $\Delta \omega = 5$
cm$^{-1}$. Dipole-dipole interactions can therefore only account for
half of the dispersion obtained in the  the  frozen phonon
calculation. We suggest that the remainder of the splitting is due to
a short-range electronic interaction. This electronic interaction
gives rise to an additional splitting, and explains why $\Delta \omega > 0$ 
at $E=1.5$~V/\AA (see Fig.~3c) even though  the dipole-dipole interaction
vanishes at this field.

To simplify the calculations we will in the following use
the dipole-dipole interaction model(Eq.~(\ref{eq:hamil})) to describe  all the
interactions, and  
 determine  field dependent   parameters $\omega_{\rm E}$ and $\gamma_{\rm E}$ by
relating the $\Gamma$-point eigenmodes of the model to the calculated
frozen phonon  values. In this way we
approximate the effect of the short-range  electronic interactions by
long-range dipole forces.
To test the accuracy of this approximation we have used the  model  to calculate  $\omega_{\rm s}$ and  $\omega_{\rm a}$ at the J and
J' point in the surface Brillouin zone. The result is shown 
in Table~1 and  the comparison with the first principles calculation is reasonable.

\begin{figure}
\begin{center}
\leavevmode
\epsfxsize=84mm
\epsffile{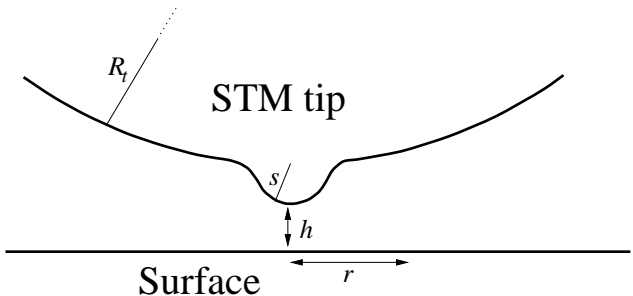}
\end{center}
\caption{The geometry used to model the electric field from the STM
tip, $R_t$ is the tip curvature, $s$
the size of the protrusion on the tip, $h$ the tip-surface distance,
and $r$ the radial distance from the tip apex. \label{fig:tipgeom}}
\end{figure}

\begin{figure}
\begin{center}
\leavevmode
\epsfxsize=84mm
\epsffile{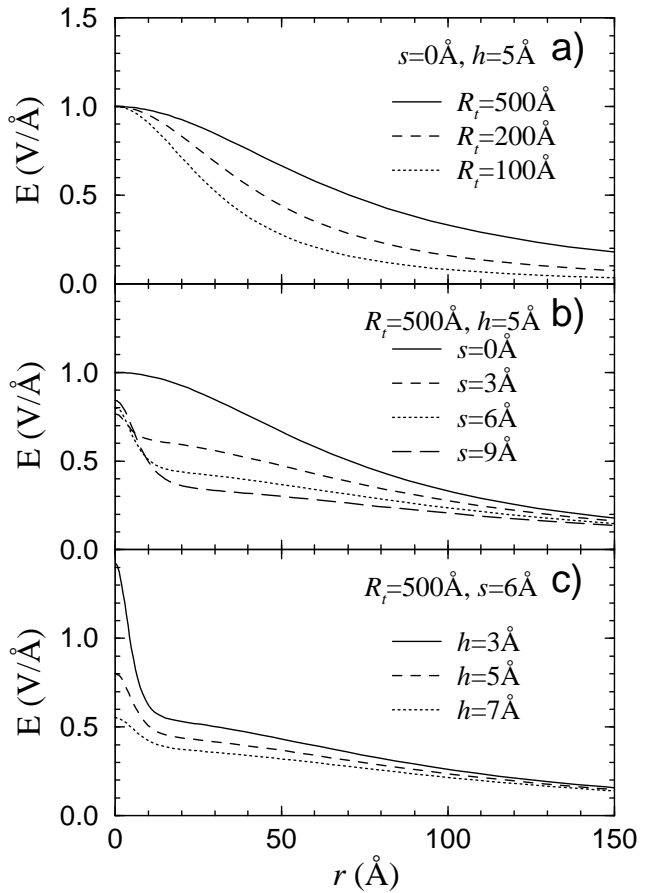}
\end{center}
\caption{Electric field at the surface for the geometry in
Fig.~\ref{fig:tipgeom} with  a potential difference of 5
V between the tip and the surface. We assume  that both the tip
and the surface are metallic. The tip parameters are: (a) s=0~\AA (no
tip protrusion), h=5~\AA , and $R_t=100$, 200, 500~\AA. (b) $R_t=500$~\AA
, h=5~\AA , and s=0, 3, 6, 9~\AA . (c) $R_t=500$~\AA , s=6~\AA , and h=3, 5, 7~\AA .\label{fig:efield}
}
\end{figure}

We next model the   electric field below the STM tip. Usually it is  found that the tip has a curvature in the range
$100$~\AA\  -- $1000$~\AA \cite{ZhIv96,OlLiYaEr98} and it is
generally accepted that the atomic resolution arises from a small
protrusion or a  single atom sticking out of the tip. We  use the geometry in
Fig.~\ref{fig:tipgeom} to model such a tip. Two parameters, the tip curvature, $R_t$, and 
the protrusion size $s$,   determine the tip geometry and we present results for
parameters in the range, $R_t=100$--$500$~\AA\ and $s=0$--9~\AA .  For
the  tip-sample distance we use, $h= 3$--$7$~\AA
, which is the typical distance range in   STM lithography experiments.
 To
find the electric field below the tip the Poisson's equation is solved
numerically using ANSYS finite element analysis\cite{ansys}. Curves in
Fig.~\ref{fig:efield}  show the radial electric field at the
surface for a potential difference of 5 V between the tip and the
surface. Curves in  Fig.~\ref{fig:efield}a show the result when there is no protrusion on the
tip ($s=0$~\AA ), and  for this geometry the electric field attains its half
value at $r \approx \sqrt{R h}$. In  Fig.~\ref{fig:efield}b results
are for a tip
with a protrusion of size $s$, and the protrusion gives rise to a reduced
electric field below the tip and it  decays 
  rapidly  around the  tip apex. The small
protrusion changes the electric field of the tip very little, and 
localization of the electric field is most pronounced for the large
protrusion. Curves in Fig.~\ref{fig:efield}c show the electric field from
the geometry with $R_t=500$~\AA\ and $s=6$~\AA\ at three tip-surface separations $h=3,6$, and $9$~\AA. The curves show that the field becomes more localized when the tip approaches the surface. 

 To determine the
vibrational states below the tip in the presence of the electric
field, we set up the Hamiltonian in Eq.~(2)
 for a finite cluster including sites up to a cutoff radius 
$r_{cut}$  and diagonalizes it numerically to find the eigenmodes
$\psi_{\alpha}$ and frequencies $\omega_{\alpha}$.  There may be
several localized modes, but we  are only interested in the localized state with the largest projection $p$ 
at the site directly below the STM tip ($r=0$). This state is determined using
\begin{equation}
p=\max_{\alpha}[ |\langle\psi_{\alpha} |0\rangle|^2 ],  
\end{equation}
where the maximum is over the eigenmodes with 
frequency outside the phonon band, $
\omega_{\alpha}-\omega_0  \not\in [-2.8V_0,9.2 V_0]$.  For the spatial electric fields
considered in this paper the value of $p$ is converged
for cluster sizes $r_{cut}=50$--$100$~\AA .

\begin{figure}
\begin{center}
\leavevmode
\epsfxsize=84mm
\epsffile{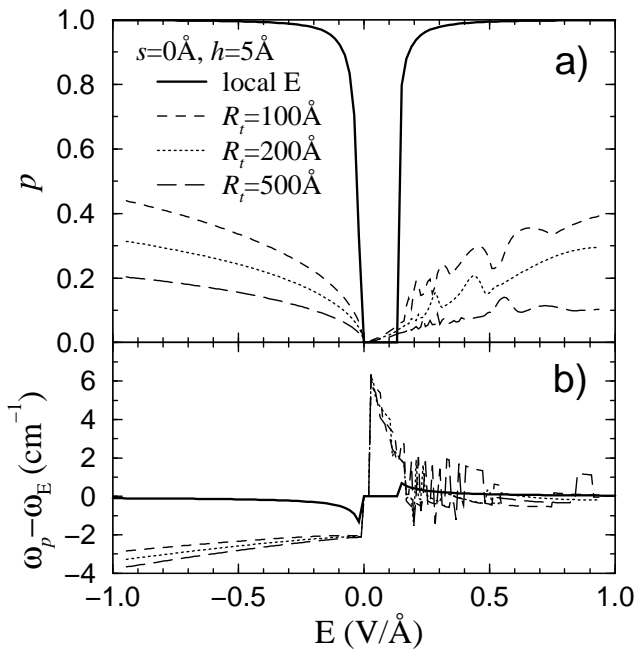}
\end{center}
\caption{a) The largest projection $p$ at site $r=0$  of the localized vibrational states when
the surface is subject to 
 the  electric fields of  Fig.~\ref{fig:efield}a)(a spherical tip with radius $R_t$), and when the field is localized at $r=0$(solid line). E is the
electric field at $r=0$. 
b) The vibrational frequency, $\omega_p$,   of the localized state  relative
to the local frequency $\omega_{\rm E}$ at site $r=0$. \label{fig:s0loc}}
\end{figure}

Curves in Fig.~\ref{fig:s0loc} show $p$ and the corresponding vibrational
frequency  $\omega_p$ when the surface is subject to  the fields of
Fig.~\ref{fig:efield}a. The ``local E'' curve corresponds to the
geometry of Persson and Avouris\cite{PeAv95} 
where the electric field is
localized at $r=0$.  In this case a localized state is split of
the phonon band at all negative fields, while at positive bias a
threshold field of $0.12$~V/\AA\ is needed to obtain localization. For typical fields in H desorption experiments, 0.5--1
V/\AA , the  state is completely localized at the site below the
tip ($p=1$) and $\omega_p$ is similar to the
frequency, $\omega_{\rm E}$, of the local oscillator at $r=0$. In the case of a
 tip with radius, $R_t$, a localized state exists for nearly all
fields, i.e. at positive bias the threshold field is 0.03~V/\AA . The
lower positive threshold field compared to the ``local E'' case is
obtained because the mode is a superposition of 
 several sites with $\omega_i \sim \omega_{\rm E}$. For typical fields 
in desorption experiments the mode has a substantial weight, $p\sim 0.3$, at
the site below the tip. 

\begin{figure}
\begin{center}
\leavevmode
\epsfxsize=84mm
\epsffile{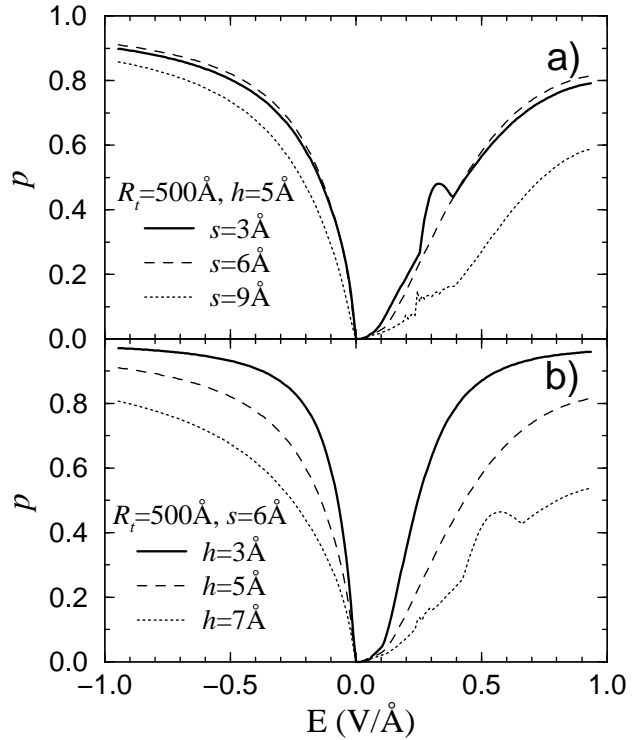}
\end{center}
\caption{Similar to Fig.~\ref{fig:s0loc}a but for the electric fields of
 (a) Fig.~\ref{fig:efield}b, and  (b)  Fig.~\ref{fig:efield}c. \label{fig:s1loc}}
\end{figure}

In Fig.~\ref{fig:s1loc} we show the effect of a small protrusion on
the STM tip. In this case  the spatial localization is improved, and for fields  0.5--1~V/\AA\ we have $p\approx
0.8$. Thus we confirm the results of Persson and Avouris\cite{PeAv95}, that there
exists a localized mode in the region below the tip, however, it is not completely
localized at a single site. 

\section{Decay of the localized vibration}
\label{sec:decay}
 Consider an STM experiment where a tunneling electron
scatters inelastically with the H atom below the tip and the H
atom is excited into the $n=1$ vibrational state of the stretch
mode. We now consider the decay of such an excitation.
There are  three important time scales, the coherent transfer time,
$\tau_{c}$, the phase relaxation time
$\tau_{ph}$ and  the energy relaxation time $\tau_{en}$. The coherent
transfer time is the time it takes for the local excitation to be
transfered into the localized eigenmode below the tip,
$\tau_{c}\sim \hbar/\Delta \omega \approx 0.5$ ps. Next the eigenmode
looses its phase  due to coupling with  a $200$~cm$^{-1}$ Si
phonon\cite{TuChRaBoLu85,GuLiHi95} and  the phase relaxation time has
been measured to be $\tau_{ph}
\approx 8$  ps\cite{TuChRaBoLu85} at room 
temperature, and  $\tau_{ph}
\approx 75$  ps\cite{GuLiHi95} at 100 K.  Finally the energy of the mode
will decay into the Si substrate via a coupling with three Si-H bending
modes(600~cm$^{-1}$) and one 300~cm$^{-1}$ Si phonon. The time scale for this process is 
$\tau_{en} \approx  10$ ns at room
temperature\cite{FoKaLyAv98,GuLiHi95}. 

In the previous section we found  a 
localized eigenmode   with  $p\sim 0.8$. 
The  excitation at $r=0$ will be a superposition of this mode and
 more extended states.   After $\tau_{c}$ the extended states
have diffused away, thus 
$ 80$\% of the initial excitation is in the localized eigenmode, and  the total 
 probability of  finding the initial excitation at $r=0$ is $p^2 \sim 0.6$.
 For $t>\tau_{ph}$ the excitation  can diffuse  away to the
neighbouring H atoms  due to dipole-dipole couplings. This is the
so-called F\"{o}rsters mechanism for incoherent diffusion, and   in the
following  we will calculate the incoherent diffusion rate, $w$, using F\"{o}rsters
formula\cite{Fo48,PeAv95}
\begin{eqnarray}
\label{eq:forster}
w&=&\frac{2}{ \pi}\sum_{i\neq 0} \int_{-\infty}^{\infty}
|H_{0i}|^2 A_0^0(\omega ) A_i^0(\omega) d\omega .
\end{eqnarray}
In this equation  $A_i^0$ is the spectral function at site $i$ for
noninteracting H modes ($H_{ij}^0\propto \delta_{ij}$), but including
the coupling with substrate phonons which gives rise to the phase relaxation.  The spectral functions are
obtained from  the noninteracting retarded Greens  functions
\begin{eqnarray}
G_i^0(t)& = &-i\Theta(t) \langle [\hat c_i(t), \hat c_i^\dagger (0)]
\rangle , \\
 A_i^0(\omega) & = & 2 {\rm Im} G_i^0(\omega),
\end{eqnarray}
where $\hat c_i^\dagger$ and  $\hat c_i$ are local creation and annihilation
operators of the stretch mode.  The phase relaxation
can be described approximately by the Hamiltonian\cite{PeHoRy86}
\begin{equation}
H_{ii}^0 = \hbar ( \omega_i+ \delta \omega \hat n_i) \hat c_i^\dagger \hat c_i,
\end{equation}
where $n_i = \hat b_i^\dagger \hat b_i$ is the projected occupation operator of
the $\Omega=200$~cm$^{-1}$ Si phonon, and $\delta \omega$ is the change
in the local  frequency when the Si phonon is excited from level $n$ to $n+1$. 
The correlation functions of $n_i$ have been calculated by Persson
{\it et al.}\cite{PeHoRy86}
\begin{eqnarray}
\langle n(t) \rangle & =&  n_{\rm B}(\Omega  ), \\
\langle n(t)n(0)\rangle &= & \langle n \rangle (1+\langle n \rangle) e^{-\eta t}+ \langle n \rangle^2.
\end{eqnarray}
 The friction parameter $\eta$ describes the damping of the
Si phonon, and $n_{\rm B}(\omega)=(e^{\beta \omega}-1)^{-1}$ is
the Bose occupation number and $\beta$ the inverse temperature.

\begin{figure}
\begin{center}
\leavevmode
\epsfxsize=84mm
\epsffile{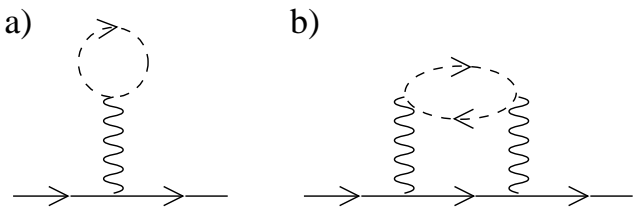}
\end{center}
\caption{Diagrams used for calculating the self energies (a)
$\Sigma_i^{(1)}$ and (b)  $\Sigma_i^{(2)}$. Solid lines symbolize propagators of
the stretch mode, while dashed lines are propagators of the Si
phonon. Wavy lines symbolize the interaction $ \delta\omega$. \label{fig:feynmann}}
\end{figure}

We now use the Matsubara formalism\cite{Ma90} to
obtain $G_{i}^0$ from an perturbation expansion in $\delta \omega$. We only consider the  two lowest
order diagrams shown in  Fig.~\ref{fig:feynmann}, and the
corresponding  self energies are 
\begin{eqnarray}
\Sigma^{(1)}_i & = & \hbar \delta \omega \langle n \rangle,\\
\Sigma^{(2)}_i & = & \hbar \delta \omega^2 \langle n \rangle (1+\langle n \rangle) \frac{1}{\omega -\omega_i+ i \eta}.
\end{eqnarray}
The $\Sigma^{(1)}$ term  gives rise to a small  frequency shift,
while the $ \Sigma^{(2)}$ leads to a damping of the mode. Considering only the
latter term, we find 
\begin{eqnarray}
A_i^0(\omega) & \approx & \frac{\Gamma_i(\omega)/\hbar}{ (\omega-\omega_i)^2 +
\Gamma_i(\omega)^2/4}, \\
\Gamma_i(\omega) &=& \frac{2 \delta \omega^2}{\eta}\frac{  n_{\rm B}(\Omega)[n_{\rm B}(\Omega)+1]}{(\omega-\omega_i)^2/\eta^2 + 1},
\end{eqnarray}
where  $\Gamma_i(\omega_i)$
is the phase relaxation rate and we have   used $\Gamma_i(\omega_i) \gg \eta$.
Thus the spectral function  resembles  a
Lorentzian with Full Width at Half Maximum(FWHM) $\Gamma_i(\omega_i)$ for $\omega \sim \omega_0$ and it decays as   $(\omega-\omega_0)^4$
in the tails. From the experimental dephasing lifetimes\cite{TuChRaBoLu85,GuLiHi95} we obtain
$\Gamma_i(\omega_i)=1/\tau_{ph}\approx0.7$~cm$^{-1}$ at room temperature. 
We  estimate the coupling strength  using  $\delta \omega_i \approx-\Omega
\omega_i/4E_{\rm d}=-4$~cm$^{-1}$\cite{PeRy89}, and the friction parameter
can  then be determined from $\eta= 2 \delta
\omega^2 n_{\rm B}(\Omega)(1+n_{\rm B}(\Omega))/2\Gamma_i(\omega_i)
\approx 50$~cm$^{-1}$. The values of $\delta \omega$ and $\eta$ obtained in this
way are similar to the measured  room temperature  values
 for Si(111)\cite{DuChHi90}.

\begin{figure}
\begin{center}
\leavevmode
\epsfxsize=84mm
\epsffile{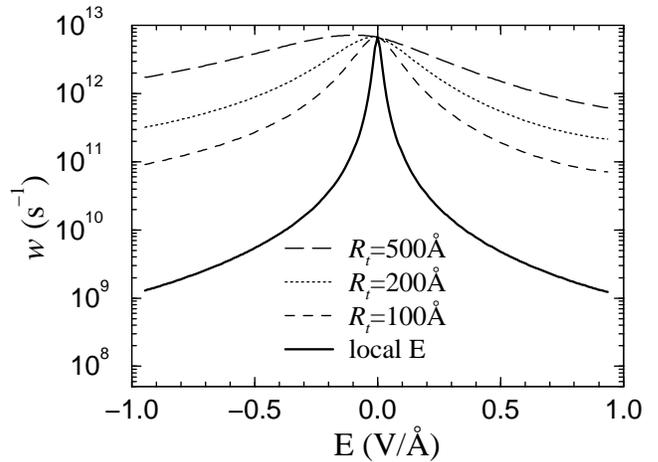}
\end{center}
\caption{Incoherent lateral diffusion rates, $w$, for the local $n=1$
vibrational excitation at $r=0$, when
the surface is subject to the electric field of a spherical tip with
radius $R_t$ at $h=5$\AA\ above the surface(broken curves), or the field is localized at $r=0$(solid). E is the
electric field at $r=0$.(see also Fig.~\ref{fig:s0loc})  \label{fig:forster0}}
\end{figure}

\begin{figure}
\begin{center}
\leavevmode
\epsfxsize=84mm
\epsffile{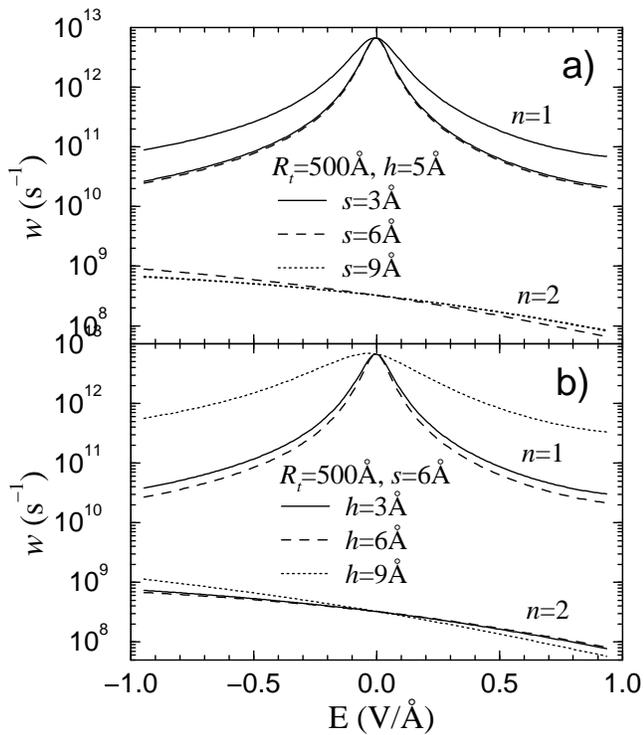}
\end{center}
\caption{Incoherent lateral diffusion rates, $w$, for the $n=1$ and
$n=2$ mode at $r=0$ as function
of the electric field of (a) geometries in Fig.~\ref{fig:efield}b, and
(b) geometries in  Fig.~\ref{fig:efield}c. \label{fig:forster}}
\end{figure}

To obtain the diffusion rate we perform the integration in
Eq.~(\ref{eq:forster}) thus obtaining
\begin{equation}
w \approx \frac{ 4}{\hbar^2}\sum_{i \neq 0} \frac{\chi_0^2 \chi_i^2}{
r_i^{6}}\frac{\Gamma_0(\omega_0)+\Gamma_0(\omega_i)}{	[\omega_i-\omega_0]^2+[\Gamma_0(\omega_0)+\Gamma_0(\omega_i)]^2/4}.
\label{eq:forsteri}
\end{equation}
In the case where $\Gamma_0(\omega_i) = \Gamma_0(\omega_0)$ this result is
similar to  that  of Ref.~\cite{PeAv95}.

 Curves in Fig.~\ref{fig:forster0} show the 
values of $w$ as obtained from Eq.~(\ref{eq:forsteri}) when the surface is subject to the same electric fields
as in Fig.~\ref{fig:s0loc}. The solid line corresponds to the electric
field model of
Persson and Avouris and similar to Ref.~\cite{PeAv95,PeAv97} we find
$w\sim 5 \times 10^9$ s$^{-1}$ for typical STM fields. The other
curves in Fig.~\ref{fig:forster0} and the curves in
Fig.~\ref{fig:forster} show that for  more realistic models of the tip
electric field  the value
of $w$ is more than one order of magnitude larger, and a typical value
in an STM experiment is $w\sim 10^{11}$ s$^{-1}$. Thus,  the $n=1$ vibrational
excitation at $r=0$   will  diffuse away very fast to the nearest
neighbour sites in contrast to the result of
Persson and Avouris\cite{PeAv97}. The reason for this is that for a
realistic STM geometry the
electric field at $r=0$ is not very different from the nearest
neighbour sites and there is a large diffusion rate into these
sites.

 For the decay of the $n>1$ excitation we have to take into account the
anharmonicity of the Si-H bond potential.
In section~\ref{sec:firstp} it was shown that the bond potential of the H atom is well
described by a Morse potential. The eigenstates of a Morse potential
is given by
\begin{equation}
 \hbar \omega(n) = E_{\rm d}\left[1- \frac{\alpha\hbar }{\sqrt{2{\rm M_H}
E_{\rm d}}}(n+\frac{1}{2})\right]^2,
\end{equation}
where $n$ takes positive integral values from zero to the greatest
value for which $n+\frac{1}{2} <\sqrt{2{\rm M_H} E_{\rm d}}/\alpha\hbar$. For the H potential  $n = 0,1,\ldots 24$ and 
$\omega(n)  = 0.129, 0.378, 0.618, 0.847, \ldots$ eV. The  anharmonicity 
is  substantial and $\omega(2)-\omega(1) =
\omega(1)-\omega(0)+U$, where $U=-0.010$ eV. The frequency of  the  $n=2$
state is  outside the phonon band, and this gives rise to a
localization of the state\cite{Gu91,LiVa92}. The diffusion rate of
this state can be estimated from Eq.~(\ref{eq:forsteri}) by using
$\omega_0+U$ for the frequency at site $0$, and the result of such a
calculation is shown by  the three
lower curves in figure~\ref{fig:forster}. The value
of  $w$ is of the same order of magnitude as the room temperature
energy relaxation rate ($\sim 10^8$ s$^{-1}$).  For $n>2$ the
relaxation rate is $\ll 10^8$ s$^{-1}$. Thus,  it is mainly the
lifetime of the $n=1$
excitation which is  affected by incoherent
diffusion. In the next section we will investigate  the effect of the
reduced lifetime of the $n=1$ excitation  on STM induced desorption.

\section{Calculation of the desorption rate}
\label{sec:des}
\begin{figure}
\begin{center}
\leavevmode
\epsfxsize=84mm
\epsffile{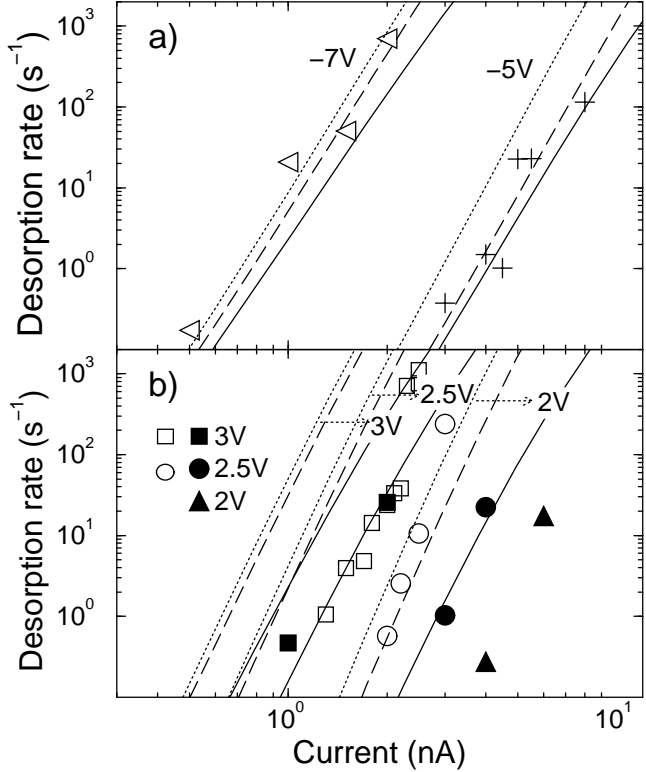}
\end{center}
\caption{Experimental desorption rate $R_t$ as a function of tunnel
current $I$ for (a) sample bias 
  $V_{b}=-7$~V(triangles) and $-5$~V(crosses),
(b) sample bias $V_{b}=2$~V(triangles), $2.5$~V(circles)  and
$3$~V(squares). Open data points are from
Ref.~\protect\cite{StThSaQuHuMuGr98,StHuThXi98} and solid data points from
Ref.~\protect\cite{ShWaAbTuLyAvWa95}. Dashed lines show the
theoretical results of Ref.~\protect\cite{StThSaQuHuMuGr98,StHuThXi98}
in which  inelastic  dipole scattering and
lateral escape of the $n=1$ vibrational state are not included. Dotted
lines show the result when   inelastic  dipole scattering is included
and solid  lines  show the full model where  both dipole scattering and 
 a lateral escape rate of $w=10^{11}$~s$^{-1}$ for the $n=1$ vibrational
state are included. \label{fig:ivrmp} }
\end{figure}

In this section we will calculate the desorption rate, $R$, of the H atom
below the STM tip, due to electron inelastic scattering through
dipole coupling or by  resonance coupling 
 with the Si-H $5\sigma$ and $6\sigma^*$ resonances.
The fraction of
electrons which scatters inelastically through
dipole coupling is given by $f_{in}^{dip}\approx (\chi_0/e
a_0)^2\approx 0.001$\cite{PeDe86}. The
theoretical model we use for calculating the inelastic current, $I_{n \rightarrow
n+N}$, due to resonance coupling has been described in
Ref.~\cite{StThSaQuHuMuGr98,StHuThXi98}. 
In those works we only
considered resonance coupling and decay through  energy relaxation  with $w_{\rm
en}=1/\tau_{\rm
en}=10^{8}$~s$^{-1}$, and dotted lines in  Fig.~\ref{fig:ivrmp}
correspond to those results. The dashed lines show the result of
including dipole coupling  and the little difference between the
dotted and dashed lines  justify the neglect of dipole coupling in our
previous studies. The solid lines  in  Fig.~\ref{fig:ivrmp}
show the result of including both dipole coupling and
lateral diffusion of the $n=1$
excitation with $w=10^{11}$~s$^{-1}$. Defining
$\mu(w)=R(w)/R(w=0)$ as the suppression of the desorption due to
lateral diffusion of the excitation, we find $\mu \sim 0.1$--$0.3$ at
negative bias and $\mu \sim 0.02$--$0.08$ at positive bias.  Using
$w=10^{10}$~s$^{-1}$ or $w=10^{12}$~s$^{-1}$ changes $\mu$  less
than 10 percent. This
is quite different from the model of Persson and Avouris\cite{PeAv97}
where $\mu \sim w/w_{\rm en}$. The reason is that in our model we include
multiple phonon excitations, i.e. we use $N=1,2,3$ in the calculation  of the
inelastic current\cite{StThSaQuHuMuGr98,StHuThXi98}.  When the
lateral diffusion rate of the $n=1$ level is large, the desorption
proceeds via a direct excitation from $n=0$ to $n=2$. At negative
biases $<-5$~V the rate of double excitations relative to single
excitations is
 $I_{n \rightarrow n+2}/I_{n \rightarrow n+1}=0.07$--$0.15 \times(n+2)
$, while  at positive biases $>2$~V it is  
$0.015$--$0.04\times(n+2)$. Thus, the larger $\mu$ at negative bias relative to
positive bias is due to a  higher probability of a multiple excitation.

\section{Summary}
\label{sec:sum}
We have studied the effect of electric field on incoherent lateral
diffusion of vibrational excitations and its implications for  STM
induced desorption of H from Si(100)-H(2$\times$1). We calculated
 the electric field at the surface for realistic STM  tip geometries and determined
the field dependent vibrational properties of the H overlayer based on  first principles 
calculations of vibrational Stark shifts and dipole-dipole interaction
matrix elements. We found that the electric field will localize the
vibrational states below the STM tip, however,  the lifetime of the $n=1$ excitations is
short ($\sim 10$ ps) due to incoherent diffusion. The diffusion of 
 higher level excitations $n>1$ is suppressed due to
anharmonic frequency shifts. 
The damping of the STM induced desorption of H due to the
lateral escape of the $n=1$ excitation  depends on the fraction of
multiple phonon excitation events 
relative to one phonon events in the inelastic scattering process. At
low positive biases we find a damping
of the desorption rate by $\mu \sim 0.02$--$0.08$, while at
negative bias   $\mu \sim 0.1$--$0.3$, reflecting the
higher probability of inelastic scattering events with a multiple phonon
excitation at the negative biases. There are no adjustable parameters
in our model and the calculated desorption rates
are in quantitative agreement with measured desorption rates.

\acknowledgements I  acknowledge Jan Tue Rasmussen
for making the ANSYS finite-element calculations, and thank 
Ben Yu-Kuang Hu, U. Quaade and F. Grey for valuable discussions and
careful reading of the manuscript. This work was 
supported by  the Danish Ministries of Industry and
Research through project No. 9800466 and the use of national computer
resources was supported by the Danish Research Councils.

\end{multicols}  

\begin{thebibliography}{10}
\bibitem{pseudo}
For Si we use 6 projectors to describe the $3s$-, $3p$- and $3d$-valence
  states, with core radii of 1.7, 1.7 and 1.9 a.u. respectively. For H we use
  two projectors to describe the $1s$ orbital and the core radius is 0.9 a.u.
  Both pseudopotentials include the nonlinear core
  correction\protect\cite{LoFrCo82}, with the densities augmented within 1.6
  a.u. for Si and 0.9 a.u. for H. Atomic transferability tests show that the
  errors in the eigenstates of the exited atoms are less than 4 meV.
\bibitem{cellsize}
For the frozen phonon calculation at the J point we use a (1$\times$4) cell,
  and at the J' point we use a (2$\times$2) cell.
\bibitem{image}
Silicon has a large dielectric constant($\sim 12\epsilon_0$) and the surface
  behaves almost as a metallic surface.
\bibitem{elastic}
The neglect of substrate mediated forces is justified by the high frequency of
  the stretch mode relative to bulk Si frequencies.
\end{thebibliography}
\end{document}